\documentclass[12pt]{iopart}
\usepackage{graphicx}
\usepackage{amssymb}
\usepackage[pdftitle={Gravitational solution to the Pioneer 10/11 anomaly}, pdfauthor={J. R. Brownstein and J. W. Moffat},
bookmarks, bookmarksopen, colorlinks=true, linkcolor=blue, citecolor=blue, anchorcolor=blue, urlcolor=blue]{hyperref}
\newcommand{\au}{\,\mbox{AU}}
\newcommand{\preprint}[1]{\href{http://arxiv.org/abs/#1}{{\it Preprint} #1}}
\newcommand{\citeasnoun}[1]{Ref.\,\cite{#1}}

\begin{document}
\title{Time Delay Predictions in a Modified Gravity Theory}
\author{J.\,W. Moffat}
\address{Perimeter Institute for Theoretical Physics, Waterloo, Ontario, N2L 2Y5, Canada}
\address{Department of Physics, University of Waterloo, Waterloo, Ontario N2L 3G1, Canada}
\ead{\href{mailto:john.moffat@utoronto.ca}{john.moffat@utoronto.ca}}

\begin{abstract}
The time delay effect for planets and spacecraft is obtained from
a fully relativistic modified gravity theory including a fifth
force skew symmetric field by fitting to the Pioneer 10/11
anomalous acceleration data. A possible detection of the predicted
time delay corrections to general relativity for the outer planets
and future spacecraft missions is considered. The time delay
correction to GR predicted by the modified gravity is consistent
with the observational limit of the Doppler tracking measurement
reported by the Cassini spacecraft on its way to Saturn, and the
correction increases to a value that could be measured for a
spacecraft approaching Neptune and Pluto
\end{abstract}

\section{Introduction}

In previous work~\cite{Brownstein}, we have provided a possible
gravitational explanation for the Pioneer 10/11 anomalous
acceleration, based on the scalar-tensor-vector gravity (STVG)
theory~\cite{Moffat}. In the following we will obtain predictions
for the corrections to the GR (general relativity) Shapiro time
delay~\cite{Shapiro,Shapiro2,Reasenberg,Bertotti,Will}, obtained
from the fitted anomalous acceleration observed in the Pioneer
10/11 spacecraft
data~\cite{Anderson,Anderson2,Turyshev,Anderson3,Turyshev2}. For
our study of the solar system, we must account in our modified
gravity theory for the variation of $G$ with respect to the radial
distance $r$ from the center of the Sun. Since we do not possess
rigorous solutions of the field equations, we use a
phenomenological parameterization of the varying parameters
$\alpha(r)$ and $\lambda(r)$ to obtain fits to the anomalous
Pioneer acceleration data that are consistent with the solar
system and fifth force experimental bounds. Thus, with the
variation of $G$ with distance from the Sun, we can make
predictions for other observational tests in the solar system. In
the following, we shall calculate the corrections to the GR time
delay effect for planets and spacecraft probes obtained from the
STVG modified gravity theory.

\section{Pioneer Anomalous Acceleration}

We assume that gravity is the cause of the Pioneer 10/11 anomaly
due to the difference between the running $G(r)$ and the bare
value, $G_{0}\sim G_N$, where $G_N$ denotes the Newtonian
gravitational constant. The Pioneer anomalous acceleration
directed towards the center of the Sun is given
by~\cite{Brownstein,Moffat}:
\begin{equation}
a_P=-\frac{\delta G(r)M_{\odot}}{r^2},
\end{equation}
where
\begin{equation} \label{deltaG}
\delta G(r)=G_N\alpha(r)\biggl[1-\exp(-r/\lambda(r))
\biggl(1+\frac{r}{\lambda(r)}\biggr)\biggr].
\end{equation}
We will use the following parametric representations of the
``running'' of $\alpha(r)$ and $\lambda(r)$~\cite{Brownstein}:
\begin{equation}
\label{alpha} \alpha(r) =\alpha_\infty(1-\exp(-r/{\bar r}))^{b/2},
\end{equation}
\begin{equation}
\label{lambda} \lambda(r)=\lambda_\infty{(1-\exp(-r/{\bar
r}))^{-b}}.
\end{equation}
Here, ${\bar r}$ is a non-running distance scale parameter and $b$
is a constant\footnote{Note that in ref.~\cite{Brownstein} the
exponent $b$ in Equation (12) should be $-b$.}.

A best fit to the acceleration data extracted from
\citeasnoun{Anderson3} has been obtained using a nonlinear
least-squares fitting routine including estimated errors from the
Doppler shift observations~\cite{Anderson2}. The best fit
parameters are~\cite{Brownstein}:
\begin{eqnarray}
\nonumber \alpha_\infty &=& (1.00\pm0.02)\times 10^{-3},\\
\nonumber \lambda_\infty &=& 47\pm 1\au ,\\
\nonumber {\bar r} &=& 4.6\pm 0.2\au ,\\
\label{bestparameters} b &=& 4.0.
\end{eqnarray}
The small uncertainties in the best fit parameters are due to the
remarkably low variance of residuals corresponding to a reduced
$\chi^{2}$ per degree of freedom of 0.42 signalling a good fit.

\section{Time Delay in Modified Gravity}

Our scalar-tensor-vector gravity (STVG) theory does not have an
exact spherically symmetric static solution. However, we have
shown that the solution will have the form of the Schwarzschild
solution for large values of the radial coordinate
$r$~\cite{Moffat}. In the derivation of the line element, we have
neglected the ``running'' of $G$. The line element in isotropic
coordinates is given by
\begin{equation}
ds^2=\frac{\biggl(1-\frac{GM}{2c^2r}\biggr)^2}
{\biggl(1+\frac{GM}{2c^2r}\biggr)^2}c^2dt^2
-\biggl(1+\frac{GM}{2c^2r}\biggr)^4d\sigma^2,
\end{equation}
where
\begin{equation}
d\sigma^2=dr^2+r^2d\theta^2+r^2\sin^2\theta d\phi^2.
\end{equation}
 Expanding in powers of $GM/r$, we obtain in
Cartesian coordinates
\begin{equation}
\label{metric}
ds^2=\biggl[1-\frac{2GM}{c^2r}+2\biggl(\frac{GM}{c^2r}\biggr)^2\biggr]c^2dt^2
-\biggl(1+\frac{2GM}{c^2r}\biggr) (dx^2+dy^2+dz^2),
\end{equation}
where $r=(x^2+y^2+z^2)^{1/2}, \theta=\arctan[z/(x^2+y^2)^{1/2}]$
and $\phi=\arctan(y/x)$. The Sun is taken to be at the origin of
coordinates and Earth and the spacecraft (planet) lie in the $z=0$
plane and the transmission null ray lies along the x direction.
For a null ray
\begin{equation}
ds^2=g_{00}c^2dt^2+g_{11}dx^2=0.
\end{equation}
We now replace $G$ in (\ref{metric}) by $G_N+\delta G$ where
$\delta G$ is given by $(\ref{deltaG})$ and obtain the correction
to the GR round-trip time delay:
\begin{equation}
\label{timedelay} \delta \tau=
\biggl(\frac{4M_{\odot}}{c^3}\biggr)\int_{-r_{\oplus}}^{r_p}dx\biggl[\frac{\delta
G((x^2+r_0^2)^{1/2})}{(x^2+r_0^2)^{1/2}}\biggr],
\end{equation}
where $r_{\oplus}$ and $r_p$ are the distances of the Sun from
Earth and the spacecraft (planet), respectively, and $r_0=y={\rm
const.}$ The correction $\delta\tau$ in (\ref{timedelay}) is
obtained by performing a numerical integration. The excess time
delay for a round-trip delay is given in GR by
\begin{equation}
(\Delta\tau)_{GR}=\frac{4G_NM_{\odot}}{c^3}
\ln\biggl(\frac{4r_pr_{\oplus}}{r_0^2}\biggr).
\end{equation}
The excess delay is calculated for a signal that grazes the limb
of the Sun, $r_0\sim R_{\odot}$, and it is a maximum when the
spacecraft (planet) is at superior conjunction.

In ref.~\cite{Brownstein}, the variation of $\delta G/G_N$ arising
from \Eref{deltaG} for the parametric values of $\alpha(r)$ and
$\lambda(r)$ of \Eref{alpha} and \Eref{lambda}, respectively, was
obtained using the best fit values for the parameters given in
\Eref{bestparameters}. The behavior of $G(r)/G_N$ is closely
constrained to unity over the inner planets until beyond the orbit
of Saturn ($r \gtrsim 10 \au$) where the deviation of $G(r)$ from
Newton's constant increases to an asymptotic value of
$G_{\infty}/G_N \rightarrow 1.001$ over a distance of hundreds of
$\au$.

In Table 1, we display the predicted values of the GR time delay
$(\Delta\tau)_{\rm GR}$, the correction $\delta\tau$ obtained from
STVG and the observational limit.

\begin{table}[ht]
\caption{Theoretical predictions for the corrections $\delta\tau$
to the GR time delays compared to the observational limits.}
\begin{indented} \item[]
\begin{tabular}{@{}lcccc}
\br
r &GR prediction & STVG prediction & Observational limit \\
(AU)&$(\Delta\tau)_{\rm GR}$\ $(\mu{\rm sec})$ &$\delta \tau(\mu{\rm sec})$
& $(\Delta \tau)_{\rm obs\, limit}$ $(\mu{\rm sec})$\\
\mr
1.52 (Mars)& \lineup\ 247 & $3.07\times 10^{-12}$ & $0.5$\\
5.20 (Jupiter)& \lineup\ 271 & $ 3.10\times 10^{-7}$ & -- \\
8.43 (Cassini)& \lineup\ 262& $8.71\times 10^{-6}$ & $6.0\times 10^{-3}$ \\
19.22 (Uranus)& \lineup\ 297 & $3.49\times 10^{-4}$ & -- \\
30.06 (Neptune)& \lineup\ 305 & $1.15\times 10^{-3}$ & -- & \\
39.52 (Pluto) & \lineup\ 311 & $2.06\times 10^{-3}$ & -- \\
100            & \lineup\ 329 & $9.32\times 10^{-3}$ & -- \\
\br
\end{tabular}
\end{indented}
\end{table}

The Cassini spacecraft while on its way to Saturn reported the
most accurate Doppler tracking measurement of the time delay
effect~\cite{Bertotti}. The accuracy of the measurement was made
possible by using both X-band (7175 MHz) and Ka-band (34316 MHz)
radar, thereby reducing significantly the dispersive effects of
the solar corona. The 2002 superior conjunction of Cassini was
favorable with the spacecraft at $8.43 \au$ from the Sun, and the
distance of closest approach of the radar signals to the Sun was
$r_0\sim 1.6\,R_{\odot}$.

We see from the result for the correction to the GR time delay for
the Cassini space probe that the prediction is well below the
observational limit of $6.0\times 10^{-3}\,\mu{\rm sec}$. The
reported value for the measured post-Newtonian parameter $\gamma$
is $\gamma-1=(2.1\pm 2.3)\times 10^{-5}$. Forthcoming new data for
the anomalous acceleration will provide a more accurate
determination of the parameters for $\alpha(r)$ and $\lambda(r)$
and the predicted values of $\delta\tau$.

A spacecraft orbiting Neptune and Pluto or on its way out of the
solar system could hopefully report a sufficiently accurate
Doppler tracking measurement of the time delay that could detect
the correction to the GR prediction. At the positions of Neptune
and Pluto and for a spacecraft at a distance from the Sun,
$r_0=100 \au$, the corrections $\delta\tau$ to the GR time delay
are of the order of the observational limit of the Cassini Doppler
tracking measurement, so Doppler measurements with comparable
accuracy reported by a spacecraft approaching Neptune and Pluto
and beyond the solar system could distinguish GR from the modified
gravity theory prediction based on the Pioneer 10/11 data.

The Doppler measurements should be carried out at conjunction with
the spacecraft (planet) to maximize the effects of the curvature
of space at the distance of closest approach to the Sun. Moreover,
Doppler measurements should be reported by the spacecraft (planet)
at opposition to determine a baseline that yields the Newtonian
time taken by a signal to reach the spacecraft (planet) and return
to Earth. This latter timing will require using the JPL ephemeris
including the modified gravity (STVG) to determine the corrected
orbital parameters of the spacecraft (planet). Subtracting the
baseline timing from the measurements at the distance of closest
approach to the Sun will result in a residual correction $\delta
t$ that can test the modified gravity theory. It should be noted
that the calculation of $\delta t$ is not significantly sensitive
to increasing values of $r_0$.

\section{Conclusions}

We have demonstrated in ref.~\cite{Brownstein} that the STVG
theory can explain the Pioneer anomalous acceleration data and
still be consistent with the accurate equivalence principle, lunar
laser ranging and satellite data for the inner solar system as
well as the outer solar system planets including Pluto at a
distance of $r= 39.52 \au =5.91\times 10^{12}$ meters. The
ephemerides for the outer planets are not as well know as the ones
for the inner planets due to their large distances from the Sun.
The orbital data for Pluto only correspond to the planet having
gone round 1/3 of its orbit. It is important that the distance
range parameter lies in the region $47 \au < \lambda(r) < \infty$
for the best fit to the Pioneer acceleration data, for the range
in the modified Yukawa correction to Newtonian gravity lies in a
distance range beyond Pluto.

Perhaps, a future deep space probe can produce data that can check
the predictions for the corrections to the time delays obtained
from the Pioneer anomaly acceleration data and from our modified
gravity theory. An analysis of anomalous acceleration data
obtained from earlier Doppler shift data retrieval will clarify in
better detail the apparent onset of the anomalous acceleration
beyond the position of Saturn's orbit.

\ack This work was supported by the Natural Sciences and
Engineering Research Council of Canada. I thank Joel Brownstein,
Martin Green, Kenneth Nordtvedt, Viktor Toth and Slava Turyshev
for helpful discussions.


\addcontentsline{toc}{section}{Acknowledgments}
\addcontentsline{toc}{section}{References}

\Bibliography{10}
\bibitem{Brownstein} Brownstein\,J.\,R. and Moffat\,J\,W. 2006 {\it Class.\,
Quant.\, Grav.} {\bf 23} 3427 -- 3436\\ (\preprint
{gr-qc/0511026})
\bibitem{Moffat} Moffat\,J.\,W. 2006 {\it JCAP}\, 0603 004 (\preprint{gr-qc/0506021})
\bibitem{Shapiro} Shapiro\,I.\,I. 1964 {\it Phys.\, Rev. Lett.} {\bf 13},
789 -- 791
\bibitem{Shapiro2} Shapiro\,I.\,I. 1999 {\it Rev.\,Mod.\,Phys.} {\bf 71}
S41-S53
\bibitem{Reasenberg} Reasenberg\,R.\,D., Shapiro\,I.\,I.,
MacNeil\,P.\,E., Goldstein\,R.\,B., Breidenthal\,J.\,P.,
Brenkle\,J.\,P., Cain\,D.\,L., Kaufman\,T.\,M., Komarek\,T.\,A.,
Zygielbaum\, A.\,I. 1979 {\it Astrophys.\, J.} {\bf 234}, L219 --
L221
\bibitem{Bertotti} Bertotti\, B., Iess\, I., and Tortora\, P. 2003 {\it
Nature} {\bf 425} 374 -- 376
\bibitem{Will} Will\,C.\,W. 2005 (\preprint{gr-qc/0510072})
\bibitem{Anderson} Anderson\,J.\,D., Laing\,P.\,A.,  Lau\,E.\,L., Liu\,A.\,S., Nieto\,M.\,M. and Turyshev\,S.\,G. 1998 \PRL {\bf 81} 2858--61 (\preprint{gr-qc/9808081})
\bibitem{Anderson2} Anderson\,J.\,D., Laing\,P.\,A.,  Lau\,E.\,L., Liu\,A.\,S., Nieto\,M.\,M. and Turyshev\,S.\,G. 2002 \PR D {\bf 65} 082004 (\preprint{gr-qc/0104064})
\bibitem{Turyshev} Turyshev\,S.\,G., Nieto M.\,M.\, and Anderson\,J.\,D. 2005 (\preprint{gr-qc/0510081})
\bibitem{Anderson3} Nieto\,M.\,M. and Anderson\,J.\,D. 2005 {\it Class.\, Quant.\, Grav.} {\bf 22} 5343 -- 5354 (\preprint{gr-qc/0507052})
\bibitem{Turyshev2} Turyshev\,S.\,G., Shao\,M. and Nordtvedt\,K.\,L. 2004 {\it Int.\,J.\,Mod.\,Phys.\,D} {\bf 13} 2035--64\\
(\preprint{gr-qc/0410044})
\endbib
\end{document}